\documentclass[usenatbib,fleqn]{mnras}
\usepackage{newtxtext,newtxmath}
\usepackage[T1]{fontenc}
\usepackage{ae,aecompl}

\usepackage{amsmath}	
\usepackage{amssymb}	

\usepackage{ctable}
\usepackage{url}
\usepackage{xspace}
\usepackage[normalem]{ulem}
\usepackage{hyperref}
\usepackage[all]{hypcap}
\usepackage{graphicx}

\def\msun{{\rm\,M_\odot}}

\def\msun{{\rm\,M_\odot}}

\newcommand{\be}{\begin{equation}}
\newcommand{\ee}{\end{equation}}

\def\h2{${\rm\,H_2}$}

\def\uas{$\mu$as}

\usepackage{color}

\title[Constraining a black hole companion of M87*]{Constraining a black hole companion for M87* through imaging by the Event Horizon Telescope}
\author[Safarzadeh, Loeb, Reid]{Mohammadtaher Safarzadeh$^{1,2}$\thanks{E-mail: msafarzadeh@cfa.harvard.edu}, Abraham Loeb$^{1}$, and Mark Reid$^{1}$ 
\\\\
$^{1}$Harvard-Smithsonian Center for Astrophysics, 60 Garden St, Cambridge, MA, 02138, USA\\
$^{2}$School of Earth and Space Exploration, Arizona State University, AZ, USA\\
}

\usepackage[all]{hypcap}

\pubyear{2019}

\begin{document}
\label{firstpage}
\pagerange{\pageref{firstpage}--\pageref{lastpage}}
\maketitle

\begin{abstract}

The Event Horizon Telescope (EHT), a global very long baseline interferometric array observing at a wavelength of 1.3 mm, detected the first image of the M87 supermassive black hole (SMBH). M87 is a giant elliptical galaxy at the center of Virgo cluster, which is expected to have formed through merging of cluster galaxies. Consequently M87* hosted mergers of black holes through dynamical friction and could have one or multiple binary companions with a low mass ratio at large separations. We show that a long-term monitoring of the M87 SMBH image over $\sim$1 year with absolute positional accuracy of 1$\approx\mu$as could detect such binary companions and exclude a large parameter space
in semi major axis ($a_0$) and mass ratio ($q$), which is currently not constrained. Moreover, the presence of the accretion disk around M87* excludes a binary companion with $a_0\approx$ of order a mili parsec, as otherwise the accretion disk would have been tidally disrupted.

\end{abstract}

\begin{keywords}
(galaxies:) quasars: supermassive black holes; stars: black holes; radio continuum: general   \end{keywords}

\section{Introduction}

The Event Horizon Telescope (EHT) collaboration, a global very long baseline interferometry array observing at a wavelength of 1.3 mm,  released the first image of a supermassive black hole (SMBH) with mass $M=(6.5 \pm 0.7) \times 10^9 \msun$ at the center of M87, a nearby elliptical galaxy at a distance of 16.8 $\pm$ 0.8 Mpc \citep{Collaboration:2019ck}. 
The observed image was shown to be consistent with the shadow of a spinning Kerr black hole \citep{Collaboration:2019ck,Collaboration:2019gi,Collaboration:2019es,Collaboration:2019kv,Collaboration:2019kf,Collaboration:2019bt}.

%

An interesting question involves the presence of a binary black hole companion to M87. 
M87 is the giant elliptical galaxy at the center of the Virgo cluster of galaxies, and such central galaxies often grow in mass through mergers of cluster galaxies \citep{Hopkins2006}. When a merger of galaxies takes place, the two SMBHs come together through dynamical friction and merge through gravitational wave (GW) emission \citep{Begelman1980,WL2003A,JB2003,Sesana2004,Mayer2007Sci,Volonteri2010,Colpi2011}. This makes it likely that we might witness a companion to the M87 black hole \citep{Gebhardt2011,Walsh2013}; the main question is at what separation and what mass ratio.

Being at the center of the galaxy cluster, M87* is constantly being fed by accretion of satellite galaxies which are most numerous at low end of the galaxy mass function \citep{Martel2014,Nipoti2018}. Therefore it is more likely for the M87* to have a binary companion such as an intermediate mass black hole (IMBH) where the IMBH is located at large separations since decay time for smaller objects is longer \citep{CMG1999,WL2003A}.

One approach to detect binary SMBH population is through their GW emission at nHz frequency. The Pulsar Timing Array (PTA) has placed an upper limit on their population through the expected stochastic GW background \citep{Lentati2015,Shannon2015,Arzoumanian2016}, and individual sources \citep{Sesana2009,Batcheldor2010,SV2010,Schutz:2015jx,Yonemaru2016PASJ}.

Though we have achieved upper limits on the presence of binary SMBH through PTA, we are not limited to GW detection to discover a binary companion to M87*. Moreover, GW detection by orbit of an IMBH at large separations around M87* is not possible by PTA. However, the presence of a binary companion would result in orbit of the SMBH around the center of mass of the system. Hence, long-term monitoring of such motion could reveal the presence of the binary companion. The monitoring of the absolute position of the SMBH probes a complementary regime in the parameter space compared to GW-based probes, as for detecting GW emission we need the companion to be close to the SMBH, while detection of relative change in absolute position larger separations are favored. Related techniques have been suggested in the literature for the case of Sgr A* \citep{Hansen:2003gm,Reid:2004tf,Broderick:2011hw}.
Moreover, if the binary companion is close to the SMBH it can tidally disrupt the accretion disk that feeds the SMBH.  

In this {\it Letter}, we study how EHT can place a stringent constraint on the presence of a binary system for M87* over long-term monitoring of the absolute position of the center of the SMBH.
The structure of the paper is as follows. In \S2 we layout the formalism for the implications of a binary system for M87*. In \S3 we present our results in what part of the parameter space could be ruled out by long-term monitoring of the 
M87 location. In \S4 we discuss the technical requirement to carry out such a precise measurement, and in \S5 we summaries and discuss the caveats.

\begin{figure*}
\resizebox{3.in}{!}{\includegraphics{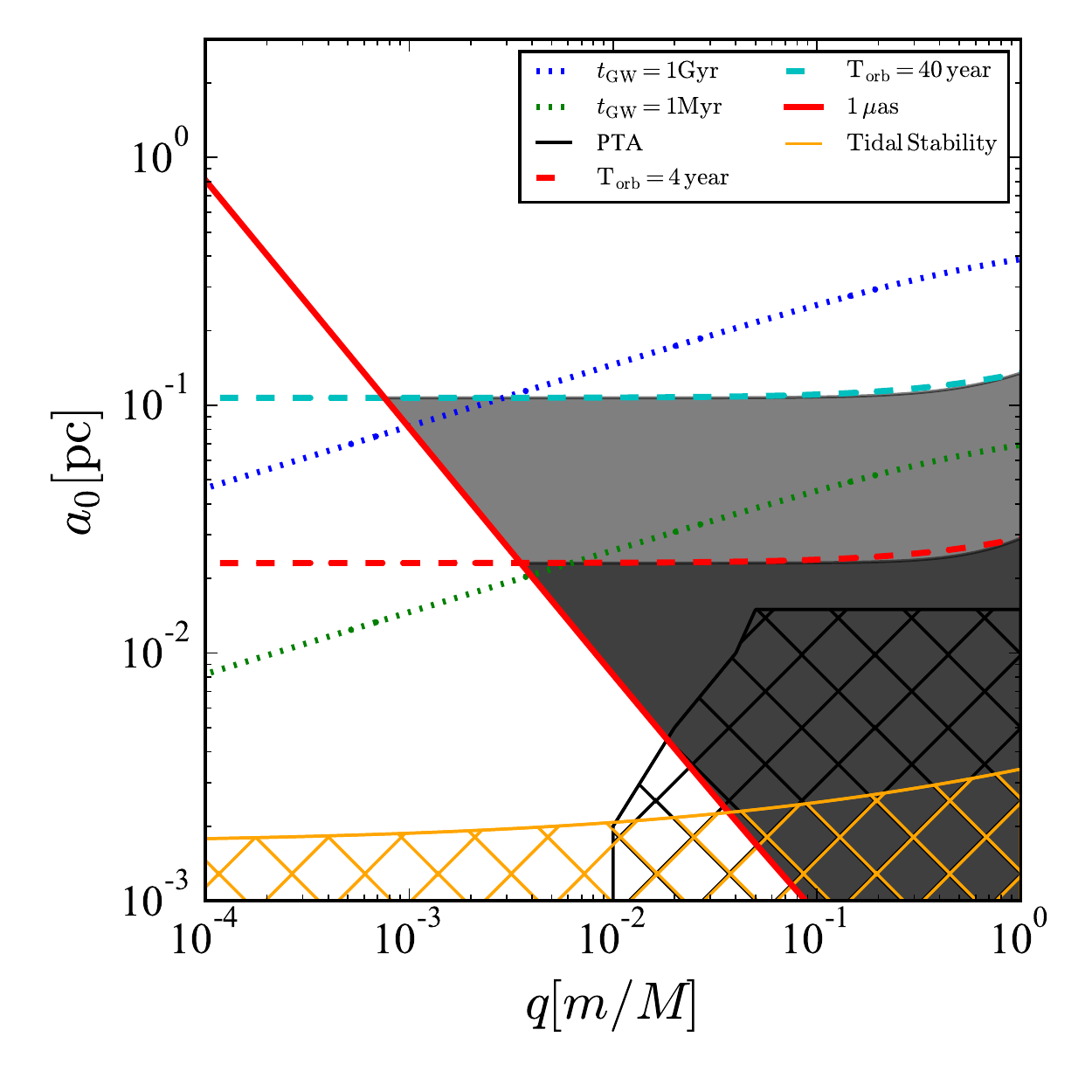}}
\resizebox{3.in}{!}{\includegraphics{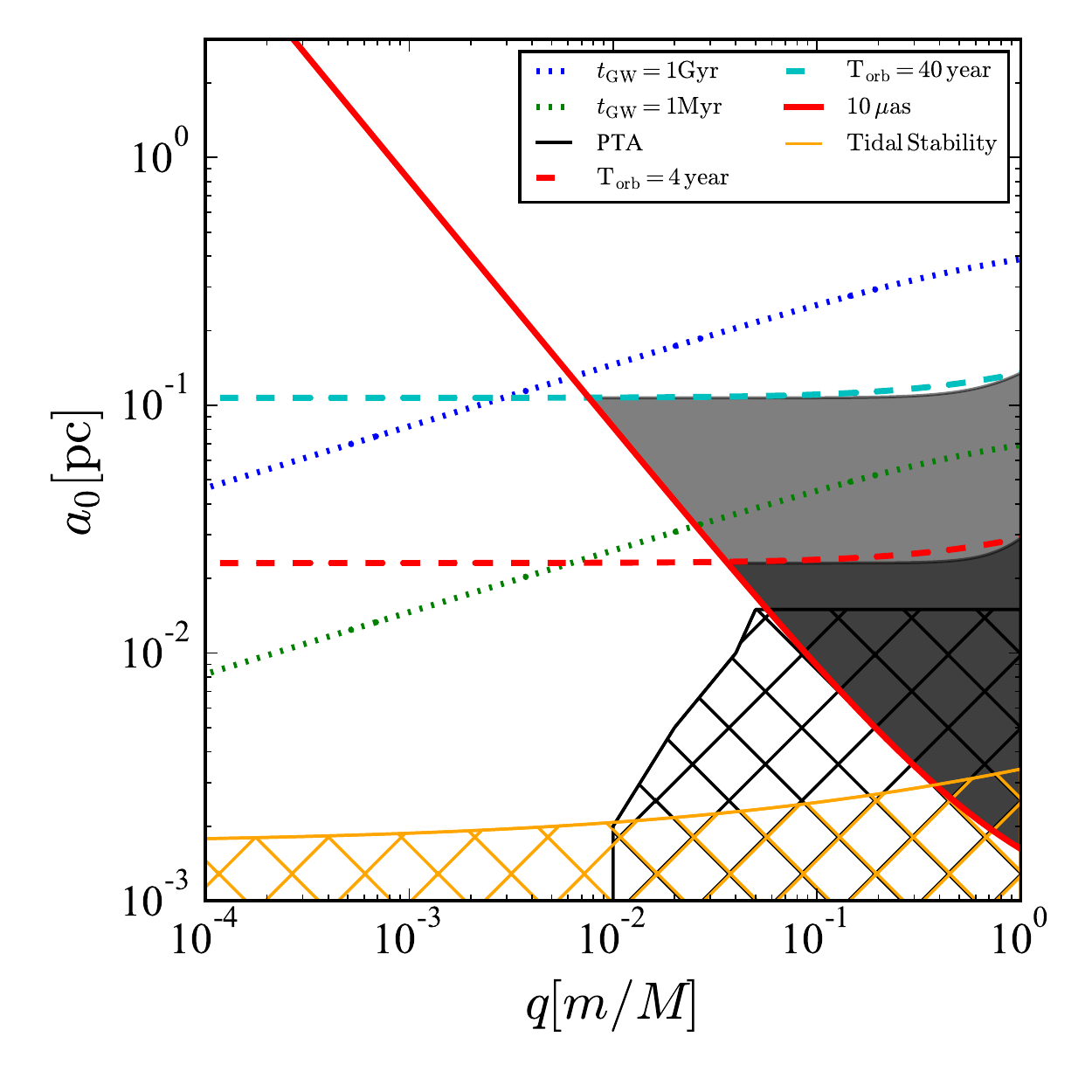}}
\caption{The excluded region of parameter space in mass-ratio versus the semi-major axis of a putative binary system for M87*. The dark (light) shaded regions correspond to monitoring the 
motion of the M87* for 1 and 10 years respectively. The solid red line shows the constraint of Eq. (\ref{eq1}). The dashed red and cyan lines show the orbital period of the system ($T_{\rm orb}$)
when $T_{\rm orb}$=4 and 40 years respectively. The dotted blue and green lines show the lifetime of the system due to the emission of GWs if the system could live for 1 Gyr and 1 Myr, respectively. 
The orange shaded region is the excluded region based on the presence of the disk shown in EHT images following Eq (\ref{eq2}). We have assumed the diameter of the disk to be $\approx$40 $\mu$as based on the EHT results. The dashed black region corresponds to the excluded region based on the PTA limit on gravitational waves \citep{Schutz:2015jx}. {\it Left panel}: assuming the positional accuracy is 1 $\mu$as. {\it Right panel}: assuming
the positional accuracy to be 10 $\mu$as. }
\label{f:fig1}
\end{figure*}

\section{impact of a binary SMBH companion for M87*}
The presence of a binary SMBH companion on M87* has two effects: (i) it makes the SMBH oscillate around the center of mass of the system, and (ii) if the binary system has small semi-major axis, then the accretion disk can get tidally disrupted. 
We denote the mass ratio of the system with $q=m/M$ where $m$ is the binary companion mass, $M$ is the mass of the M87* (6.5 $\times 10^9 \msun$). We denote the semi-major axis of the binary with $a_0$.
In the presence of a binary companion, the SMBH will move around the center of mass with an amplitude of the orbit being 
\be\label{eq1}
a= \frac{q}{1+q} \times a_0.
\ee

The orbital period associated with such an orbit is estimated through Kepler's law. We note that the maximal length of the excursion executed by the SMBH is independent of viewing angle.

Separately, an accretion disk would be tidally disrupted in the presence of the companion. An orbiting BH with mass $m$ would tidally disrupt an accretion disk with radius $r$ if $a_0$ of the binary system is less than $a_t$, where $a_t$ is given by:
\be\label{eq2}
a_t=r (1+q^{1/3})
\ee
The light ring around the silhouette of the M87 SMBH reflects the emissivity of the accretion disk and jet near the Innermost Stable Circular Orbit (ISCO). 
Therefore, we relate a major distortion in the image to a tidal disruption of the disk at a few Schwarzschild radii.
\section{results}

Figure \ref{f:fig1} shows which part of the parameter space could be excluded with long-term observation of the center of the M87*. 
On the left (right) panel, we assume the accuracy with which the motion of the M87* could be tracked is 1 (10) $\mu$as. The dark (light) shaded regions in each panel correspond to monitoring the 
motion of the M87* for 1 and 10 years respectively. The solid red lines show the limit obtained from Eq. (\ref{eq1}). The dashed red and cyan lines show the orbital period of the system ($T_{\rm orb}$)
when $T_{\rm orb}$=4 and 40 years, respectively. To track the motion of an SMBH in a binary system with orbital period $T_{\rm orb}$, a monitoring time of $T_{\rm orb}/4$ would suffice, and therefore the intersection of the solid red line with the dashed red and cyan lines translates into the constraint that would be obtained by 1 and 10 years of observations. The orange shaded region is the excluded region based on the presence of the disk shown in EHT images. If there is a binary companion at mili-parsec separation, the disk would have been tidally disrupted and the limits are obtained based on Eq. (\ref{eq2}). We have assumed the diameter of the innermost position of the disk to be $\approx$40 $\mu$as based on the EHT data.

The dotted blue and green lines delineate the lifetime constraint on the system due to the emission of gravitational waves, if the system lives for 1 Gyr and 1 Myr, respectively. The dashed black region corresponds to the excluded region based on the Pulsar Timing Array limit on gravitational waves \citep{Schutz:2015jx}.

Approaches based on detection of GWs such as PTA would require a close binary companion to be able to detect the binary SMBH. However, the technique base on the absolute position of the SMBH is sensitive to the opposite end of binary separation, meaning larger separation are favored, and therefore a larger parameter space is accessible to probe.
Our results show how a large region in parameter space could be excluded with one year of monitoring the absolute position of M87* with respect to an external background source. Moreover, the excluded region is more sensitive to the positional accuracy than the monitoring time. In other words, the difference between one year of monitoring time with 1 $\mu$as or 10 $\mu$as is more significant than the difference between monitoring for one or ten years. Therefore, it is worth exploring how much positional accuracy would be achieved with EHT.

\section{obtaining positional accuracy}
Although EHT's resolution of $25\mu$as for M87* is impressive, the imaging is done through self-calibration which loses absolute position information.  Such images made at different epochs could appear precisely registered, even if the source as a whole moves significantly.  Alternatively, calibrating interferometer phase on a different source, such as a background quasar, retains relative position information and could be used to detect the motion of M87* owing to a massive companion.  However, such phase-referencing requires switching between the calibrator and target within the interferometer coherence time which is often limited by delay fluctuations imposed by irregularities in the water vapor (e.g., clouds) above each antenna.  At the short radio wavelengths of $\approx1$ mm needed to image the inner regions of M87*, coherence times are $\sim10$ seconds, which precludes nodding an antenna to do the calibration since slewing and settling time would preclude sufficient on-source time.  However, \citet{Broderick:2011hw} suggested an intriguing method to achieve phase-referenced observation when using antenna arrays, instead of single antennas, at a site.  If one divides the array into two sub-arrays, one can simultaneously observe a background source and the target, avoiding the temporal coherence problem.  This technique does not remove spatial coherence issues (i.e., interferometer phase differences as a function of position, not time), which are analagous to the "isoplanatic patch" size for optical and infrared observations.  The analysis in \citet{Broderick:2011hw} suggests that, for source separations of $\sim1^\circ$, phase-referencing can work when observing at 230 GHz from the high, dry sites used by the EHT.

With phase-referenced data, relative positional accuracy is limited by uncompensated interferometer path-delays from large, slowly varying changes in water vapor above each telescope.   At typical sites for cm-wave VLBI (often near sea level), uncompensated path-delays are $\sim1$ cm, which lead to relative position errors of $\sim10$ \uas\ for sources separated by $\approx1^\circ$ \citep{Reid2009}. Assuming the greatly reduced water vapor experienced at the high, dry sites of EHT telescopes translates to an order-of-magnitude improvement in uncompensated path-delays compared to cm-wave VLBI sites, one could expect $\sim1$ \uas\ uncertainty in relative position measurements for sources separated by $\approx1^\circ$.  Of course, other factors may limit relative position uncertainty, including structural changes in the calibration source image.   Since such structural changes should be independent from among sources, it is wise to use more than one calibration source in astrometric observations.  \citet{Broderick:2011hw} discuss three potential calibration sources for M87* which are separated from M87* on the sky by between 1.2 and 1.8 degrees.
We note that this technique could only be used for arrays at ALMA, NOEMA, and SMA.

We note that the source in our case is a ring centroid and not a compact unresolved brightness peak of a radio source. The peculiar structure of the source likely introduces additional source of position uncertainty if phase referencing is applied to the image. However, given that we are interested in the center of the image, although the image itself is degraded by phase referencing, the ring-like structure can on the other hand increase the ability to find the center. These two competing effects need to be simulated to make a further definite statements.

\section{Summary}
SMBHs grow through the merger of black holes in the center of clusters of galaxies. M87 is the giant elliptical galaxy at the center of the Virgo cluster expected to have formed through major mergers and accretion of satellites hosting IMBHs. This would imply the possibility of one or many companion IMBHs  orbiting M87*. 

A binary companion to M87* could be detectable through emission of GW if the separations are small enough to be detectable by PTA. However, the accretion process of small satellites that host IMBHs would place their IMBH at large separations since the decay time for smaller objects is longer \citep{CMG1999,WL2003A}. This is not a favorable situation for PTA. However, we have shown that presence of a small mass companion at large separations would translate into detectable motion of the SMBH, which is achievable by EHT monitoring of the center of M87* with $\sim1$ \uas\ positional accuracy over the course of 1 to 10 years. Such positional accuracy would be achievable through relative position measurement for sources separated by $\approx1^\circ$ \citep{Reid2009}.
This regime of low mass companion at large separations is the opposite regime of that available to PTA and only could be probed by a long-term monitoring of the motion of the SMBH by EHT.

Moreover, we have shown that the presence of the accretion disk around M87* can be used to exclude the presence of a binary companion with mili-parsec separations as otherwise the accretion disk would be tidally disrupted. The limits based on tidal disruption argument are similar to those obtained through PTA.

\section{Acknowledgements}

We are thankful to Michael D. Johnson for insightful comments. 
This work was supported by the National Science Foundation under grant AST14-07835 and by NASA under theory grant NNX15AK82G as well as a JTF grant. 
MTS is grateful to the Harvard-Smithsonian Center for Astrophysics for hospitality during the course of this work. This work was supported in part by the Black Hole Initiative at Harvard University, which is funded by JTF grant.

\bibliographystyle{mnras}
\bibliography{the_entire_lib}
\end{document}